\documentstyle[12pt]{article}

\advance \textheight by \topskip
\makeatletter
\@addtoreset{equation}{section}

\makeatother

\begin{document}

\begin{titlepage}
\hbox to \hsize{\hfil TUM--HEP--272/97}
\hbox to \hsize{\hfil NF/DF--04/97}
\hbox to \hsize{\hfil April 1997}
\vfill
\large \bf
\begin{center}
Pseudoclassical model for topologically massive gauge fields
\end{center}
\vskip1cm
\normalsize
\begin{center}
Khazret S. Nirov${}^{a,}${}{\footnote{Alexander von Humboldt fellow;
on leave from the {\it Institute for Nuclear Research, Moscow, 
Russia}}${}^,$\footnote{E--mail: nirov@dirac.physik.uni-bonn.de}} $\,$ 
~and~ Mikhail S. Plyushchay${}^{b,c,}${}\footnote{E--mail: 
  plyushchay@mx.ihep.su}\\
{\small \it ${}^a{}$Institut f\"ur Theoretische Physik T30,
 Physik Department,}\\
{\small \it Technische Universit\"at M\"unchen,
 D-85747 Garching, Germany}\\
{\small \it ${}^b{}$Departamento de Fisica --- ICE, 
 Universidade Federal de Juiz de Fora}\\
{\small \it 36036-330 Juiz de Fora, MG Brazil}\\
{\small \it ${}^{c}{}$Institute for High Energy Physics, 
  Protvino, Moscow region, 142284 Russia}
\end{center}
\vskip2cm
\begin{abstract}
\noindent 
A pseudoclassical model for $P$,$T$-invariant system 
of topologically massive U(1) gauge fields is analyzed.  
The model demonstrates a nontrivial relationship between 
continuous and discrete symmetries and reveals a phenomenon 
of ``classical quantization''. It allows one to identify SU(1,1) 
symmetry and S(2,1) supersymmetry as hidden symmetries of the 
corresponding quantum system. We show this $P$,$T$-invariant 
quantum system realizes an irreducible representation of a 
non-standard super-extension of the $(2+1)$-dimensional 
Poincar\'e group.
\end{abstract}
\vfill
\end{titlepage}

\hspace{3cm}

\section{Introduction}

Revealing new symmetries and investigating their structures is the most 
powerful and productive approach in modern physics \cite{T}. Classical 
particle models are useful for finding hidden properties of corresponding 
quantum systems and understanding their nature: efficiency of the approach, 
based on pseudoclassical mechanics \cite{pm} underlying the path-integral 
formulation of the quantum field theory with fermions \cite{Ber}, was 
recently demonstrated in the context of constructing the covariant 
formulation of the superstring theory \cite{GS}, for which 
the Brink-Schwarz superparticle model \cite{BS} played very 
important role. In a previous paper \cite{NP1} we proposed a 
new pseudoclassical model, whose quantum analog is parity and 
time-reversal conserving system of topologically massive vector 
U(1) gauge fields \cite{top}. The latter may be relevant to the 
quantum Hall effect \cite{hall} and high-temperature 
superconductivity \cite{T-sc}, that basically motivates 
our studies. 

\vskip2mm
Here we continue investigations started with Ref. \cite{NP1}: 
The purpose of this paper is to uncover hidden symmetries of 
the Chern-Simons fields' system by means of the corresponding 
particle model. To be self-contained, we will recollect properties 
of our pseudoclassical construction. It contains a $c$-number 
parameter $q$, entering a mass term for spin variables, which 
displays a quantization property both at the classical and the 
quantum levels. The parameter does not affect discrete $P$ and 
$T$ symmetries of the classical theory, but turns out to be 
crucial for continuous global symmetries: there are special 
discrete values of $q$ at which the pseudoclassical model has 
a maximal set of integrals of motion. The same values of the 
parameter are separated by the requirement of maximality of global 
symmetry of the physical state space at the quantum level. Moreover, 
we shall see that only at these special values discrete symmetries 
of the pseudoclassical model are conserved upon quantization. 
It is interesting to note here that Deser, Jackiw and Templeton 
discovered the quantization of the dimensionless mass-coupling-constant 
ratio in the non-Abelian vector case having implicated symmetry arguments
\cite{top}.  

\vskip2mm
Analyzing algebras of the integrals of motion we shall reveal hidden 
SU(1,1) symmetry and S(2,1) supersymmetry of the $P,T$-invariant system 
of topologically massive vector U(1) gauge fields and show that the 
system realizes an irreducible representation of a non-standard 
super-extension of the (2+1)-dimensional Poincar\'e group. 
Though here we deal with a free theory, these results can hopefully 
find further development and applications in elementary particle and 
condensed matter physics. In this respect, it is worthwhile remembering 
that revealing of SU(N) symmetries in hadron and nuclear physics 
\cite{GN-AI} and the discovery of supersymmetry \cite{R} were realized 
first in contexts of free theories but subsequently turned out to be 
forming the basis for the description of real world. Note also that 
the first 4D group-theoretical supersymmetry constructions were 
motivated by certain aspects of parity in quantum field theory 
\cite{GL}.

\section{The field system and pseudoclassical model}

The source-free equations for topologically massive vector U(1) gauge 
fields, given in terms of a self-dual free massive field theory 
\cite{dual}, are first order differential equations 
${\cal L}^\epsilon_{\mu\nu} {\cal F}^\nu_\epsilon = 0$,
where 
${\cal L}^\epsilon_{\mu\nu} 
\equiv ( i \varepsilon_{\mu\nu\lambda} 
P^\lambda + \epsilon m \eta_{\mu\nu} )$, 
$P_\mu = -i \partial_\mu$,
$\eta_{\mu\nu} = diag(-1,+1,+1)$, 
$\epsilon = +$ or $-$, 
and we normalize the totally antisymmetric tensor 
$\varepsilon^{\mu\nu\lambda}$ by 
$\varepsilon^{012} = 1$. 
Due to the basic equations, the field 
${\cal F}^\mu_\epsilon$ 
satisfies also Klein-Gordon equation
$(P^2 + m^2) {\cal F}^\mu_\epsilon = 0$ 
and the transversality condition
$P_\mu {\cal F}^\mu_\epsilon = 0$. 
As a consequence, it carries massive irreducible representation 
of spin $s = -\epsilon 1$ of the 3D Poincar\'e group. Already 
in pioneering works \cite{top} it was noted that topological 
mass terms are parity and time-reversal odd, and the full set 
of discrete $C$, $P$ and $T$ symmetries may be restored if one 
doubles the number of fields and introduces opposite sign mass 
terms. In the case under consideration, when taking the action 
\[
{\cal A} = \int d^3x \left( 
{\cal F}^\mu_+ {\cal L}^+_{\mu\nu} {\cal F}^\nu_+ 
+ {\cal F}^\mu_- {\cal L}^-_{\mu\nu} {\cal F}^\nu_- \right),
\] 
we get $P$,$T$-invariant system of topologically massive vector 
U(1) gauge fields \cite{top}. This observation plays an essential 
role in constructing models of high-temperature superconductors.
Actually, single spin state models predict observable parity and
time-reversal violation in corresponding superconductors \cite{T-inv}, 
for which one still has no experimental evidence \cite{exp}. Besides, 
the problem of cancellation between single bare and radiatively 
generated Chern-Simons terms arises in the conventional models 
\cite{cancel}. Therefore, it is desirable to have parity and 
time-reversal conserving system modeling high-$T_c$ superconductors 
without these serious obstructions \cite{T-inv-mod}. And for our work 
it means a signal to pay particular attention to the requirement of 
$P$,$T$ invariance.

\vskip2mm
The pseudoclassical model in question here is given by the Lagrangian
\begin{equation} 
L_q = \frac{1}{2e}\left(\dot{x}_\mu 
- \frac{i}{2} v \varepsilon_{\mu\nu\lambda}
\xi^\nu_a \xi^\lambda_a \right)^2 - \frac{1}{2}e m^2 
- i q m v \xi_1^\mu\xi_{2\mu}
+ \frac{i}{2} \xi_a^\mu \dot{\xi}_{a\mu}, 
\label{L}
\end{equation}
where $x_\mu$, $\mu = 0,1,2$, denote space-time coordinates of 
the particle, $\xi^\mu_a$, $a = 1,2$, are real Grassmann variables 
forming two Lorentz vectors, $e$ and $v$ are even Lagrange multipliers, 
and $q$ is a real $c$-number parameter. 
The Lagrangian (\ref{L}) is invariant with respect to the discrete 
parity and time-reversal transformations
\[
P : X^\mu \rightarrow \tilde{\varepsilon}(X^0, -X^1, X^2),
\quad
T : X^\mu \rightarrow \tilde{\varepsilon}(-X^0, X^1, X^2),
\quad
P :, T : (e,v) \rightarrow (e,-v), 
\]
where
$X^\mu = x^\mu, \xi_1^\mu, \xi_2^\mu$, 
$\tilde{\varepsilon}=+1$ for $x^\mu$ and $\xi_1^\mu$
and $\tilde{\varepsilon}=-1$ for $\xi_2^\mu$
implying that $\xi^\mu_2$ is a pseudovector.
We have to stress that in the classical theory parity 
and time-reversal invariance take place for any value of the 
parameter $q$. Nevertheless, we shall see that the case of 
$|q| = 2$ is particular both at the classical and the quantum 
levels of the theory, and that the quantization of the model 
(\ref{L}) results in the $P$,$T$-invariant system of topologically 
massive vector U(1) gauge fields. 

\vskip2mm
To find the corresponding quantum system and reveal its hidden
symmetries, let us construct the Hamiltonian description of the
model. The nontrivial canonical
brackets following from the Lagrangian (\ref{L}) are
$\{x_\mu,p_\nu\} = \eta_{\mu\nu}$, $\{\xi_a^\mu,\xi_b^\nu\} = -
i \delta_{ab} \eta_{\mu\nu}$, $\{e,p_e\} = 1$, $\{v,p_v\} = 1$,
and we obtain two sets of primary, $p_e \approx 0$, 
$p_v \approx 0$, and secondary, 
\[
\phi = \frac{1}{2} (p^2 + m^2) \approx 0, \qquad
\chi = \frac{i}{2} \left(\varepsilon_{\mu\nu\lambda} p^\mu
\xi^\nu_a \xi^\lambda_a + 2 q m \xi_1\xi_2\right) \approx 0,
\]
constraints forming the trivial algebra of the first class 
with respect to the above brackets. As a consequence of 
the reparametrization invariance, the Hamiltonian of 
our model is a linear combination of the constraints,
$H = e\phi + v\chi + \omega_1 p_e + \omega_2 p_v$, 
with the coefficients at the primary constraints being 
arbitrary functions of the evolution parameter $\tau$. 
{}From the equations of motion,
\[ 
\dot{p}_\mu = 0, \qquad 
\dot{x}_\mu = e p_\mu + \frac{i}{2}
v \varepsilon_{\mu\nu\lambda} \xi^\nu_a \xi^\lambda_a, 
\qquad 
\dot{\xi}_{a\mu} 
= - v (\varepsilon_{\mu\nu\lambda} p^\nu \xi^\lambda_a 
- q m \epsilon_{ab} \xi_{b\mu}),
\] 
where 
$\epsilon_{ab} = - \epsilon_{ba}$, $\epsilon_{12} = 1$, 
we see that the energy-momentum vector $p_\mu$
and the total angular momentum vector 
${\cal J}_\mu = -\varepsilon_{\mu\nu\lambda} x^\nu p^\lambda 
+ \frac{i}{2} \varepsilon_{\mu\nu\lambda} \xi^\nu_a \xi^\lambda_a$ 
are integrals of motion. 

\vskip2mm
To solve the equations for the spin variables $\xi^\mu_a$, 
it is convenient to use complex mutually conjugate odd variables
$b^{\pm}_\mu = \frac{1}{\sqrt{2}} (\xi_{1\mu} \pm i \xi_{2\mu})$
with nontrivial brackets
$\{b^+_\mu,b^-_\nu\} = - i\eta_{\mu\nu}$.
Taking into account the mass-shell constraint, 
we introduce the general notation 
$f^{(\alpha)} \equiv f^\mu e^{(\alpha)}_\mu$ 
for the projection of any Lorentz vector $f^\mu$ 
onto the complete oriented triad
$e^{(\alpha)}_\mu(p)$, $\alpha = 0,1,2$, 
defined by the relations
\[
e^{(0)}_\mu = p_\mu/\sqrt{-p^2}, 
\qquad 
e^{(\alpha)}_\mu \eta_{\alpha\beta} e^{(\beta)}_\nu = \eta_{\mu\nu},
\qquad
\varepsilon_{\mu\nu\lambda} e^{(0)\mu} e^{(i)\nu} e^{(j)\lambda} =
\varepsilon^{0ij}, \quad i,j = 1,2.
\]
In terms of these, we find that the odd spin variables have 
the following evolution law:
\[
b^{(0)\pm}(\tau) = e^{\mp i q \omega(\tau)} b^{(0)\pm}(0),
\qquad 
b^{(i)\pm}(\tau) = e^{\mp i q \omega(\tau)} 
\left[ \cos\omega(\tau) b^{(i)\pm}(0) 
+ \varepsilon^{0ij} \sin\omega(\tau) b^{(j)\pm}(0) \right], 
\]
with 
$\omega(\tau) = m \int_0^\tau v(\tau^\prime) d\tau^\prime$. 
{}From these solutions we immediately obtain quadratic nilpotent 
integrals of motion: 
\[
{\cal N}_0 = b^{(0)+} b^{(0)-},
\quad 
{\cal N}_\perp = b^{(1)+} b^{(1)-} + b^{(2)+} b^{(2)-},
\quad
{\cal S} = i (b^{(1)+} b^{(2)-} - b^{(2)+} b^{(1)-}) 
\equiv {\cal J}^{(0)}.
\]

\vskip2mm
The case of $q=0$ is dynamically degenerated with the
variables $b^{(0)\pm}$ being trivial integrals of motion,
$b^{(0)\pm}(\tau) = b^{(0)\pm}(0)$. As we shall see, this 
special case is completely excluded on the quantum level.
Further, we have the nilpotent second order quantities
$B^\pm = 
\left( b^{(2)+} b^{(2)-} - b^{(1)+} b^{(1)-} \right) \pm
i \left( b^{(2)+} b^{(1)-} + b^{(1)+} b^{(2)-} \right)$
satisfying a simple evolution law: 
$B^\pm(\tau) = e^{\pm 2i\omega(\tau)} B^\pm(0)$.
We obtain that if and only if $|q| = 2$, there are two 
additional third order nilpotent integrals of motion 
in the model, namely
\[
{\cal B}^\pm_+ = B^\pm b^{(0)\pm}, 
\quad
{\cal B}^+_+ = ({\cal B}^-_+)^* \;\;
{\rm for}\; q = 2, \quad {\rm or}
\quad
{\cal B}^\pm_- = B^\pm b^{(0)\mp},
\quad 
{\cal B}^+_- = ({\cal B}^-_-)^* \;\;
{\rm for}\; q = - 2,
\]
which are local in the evolution parameter $\tau$ 
quantities.

\vskip2mm
Thus, here we have observed some phenomenon of 
{\it classical quantization}: there are two special 
values of the parameter $q$, $q = \pm 2$, when, and
only when, the system has additional (local in $\tau$) 
nontrivial integrals of motion. These integrals are the 
generators of corresponding global symmetry transformations, 
and so, the system has maximal global symmetry at these two 
special values of the model parameter.

\section{Quantization of the model}

To quantize the model, we completely remove Lagrange multipliers and 
their conjugate momenta from the theory by gauge-fixing conditions 
$e - e_0 \approx 0$, $v - v_0 \approx 0$ 
for the primary constraints, where $e_0$ and $v_0$ 
are some constants. Upon quantization, the odd variables
$b^\pm_\mu$ become the fermionic creation-annihilation operators
$\widehat{b}^\pm_\mu$ having the only nonzero anticommutators 
$[ \widehat{b}^-_\mu,\widehat{b}^+_\nu ]_+ = \eta_{\mu\nu}$.  
Then the arbitrary quantum state can be realized over the vacuum 
$|0\rangle$, $\widehat{b}^-_\mu |0\rangle = 0$,
$\langle 0|0 \rangle = 1$:
\[
\Psi(x) = \left( f(x) + {\cal F}^\mu(x) \widehat{b}_\mu^+ 
+ \frac{1}{2!} \varepsilon_{\mu\nu\lambda} {\tilde{\cal F}}^\mu(x) 
\widehat{b}^{+\nu} \widehat{b}^{+\lambda} 
+ \frac{1}{3!} \tilde{f}(x) \varepsilon_{\mu\nu\lambda}
\widehat{b}^{+\mu} \widehat{b}^{+\nu} \widehat{b}^{+\lambda} \right) 
|0\rangle.
\] 
The coefficients of this expansion are some square-integrable functions 
of the space-time coordinates. The quantum parity and time-reversal 
transformations are generated by the antiunitary operators
\[
U_P = V^0_+ V^1_- V^2_+,
\quad
U_T = V^0_- V^1_+ V^2_+,
\quad
{\rm with}
\quad 
V^\mu_\pm = \widehat{b}^{+\mu} \pm \widehat{b}^{-\mu},
\]
as follows,
\[ 
P,T: \Psi(x) \rightarrow \Psi^\prime(x^\prime_{P,T}) 
= U_{P,T} \Psi(x), 
\qquad
x^{\prime\mu}_P = (x^0,-x^1,x^2), 
\quad
x^{\prime\mu}_T = (-x^0,x^1,x^2). 
\]
In correspondence with classical relations, we have
\[
U_P \widehat{b}^\pm_{0,2} U_P^{-1} = \widehat{b}^\mp_{0,2}, \quad
U_P \widehat{b}^\pm_1 U_P^{-1} = - \widehat{b}^\mp_1, \qquad
U_T \widehat{b}^\pm_{1,2} U_T^{-1} = \widehat{b}^\mp_{1,2}, \quad
U_T \widehat{b}^\pm_0 U_T^{-1} = - \widehat{b}^\mp_0.
\]
While acting on the general state $\Psi(x)$ these operators 
induce mutual transformation of scalar, 
$f(x) \leftrightarrow \tilde{f}(x)$, 
and vector, 
${\cal F}^\mu(x) \leftrightarrow \tilde{\cal F}^\mu(x)$,
fields.

\vskip2mm
The physical states should be singled out by the quantum 
analogs of the remaining first class constraints, 
$(P^2 + m^2) \Psi = 0$ 
and
$\widehat{\chi} \Psi = 0$, 
where we assume that 
$P_\mu = -i \partial_\mu$. 
Note that the nilpotent constraint
$\chi$ admits no, even local, gauge condition, and so, the 
respective sector of the phase space can be quantized by the 
Dirac method only \cite{PR}. Let us fix in the quantum operator 
$\widehat{\chi}$ the same ordering as in the corresponding classical 
constraint. This gives 
\[
\widehat{\chi} = i\varepsilon_{\mu\nu\lambda} P^\mu 
\widehat{b}^{+\nu} \widehat{b}^{-\lambda} 
- q m ( \widehat{b}^+_\mu \widehat{b}^{-\mu} - 3/2).
\]
As a consequence of the quantum constraints, we find that 
$f(x) = \tilde{f}(x) = 0$, 
whereas for the fields 
${\cal F}_\mu(x)$ and ${\tilde{\cal F}}_\mu(x)$
we get the equations  
\begin{equation}
i \varepsilon_{\mu\nu\lambda} P^\nu {\cal F}^\lambda 
- \frac{1}{2} q m {\cal F}_\mu = 0, \qquad
i \varepsilon_{\mu\nu\lambda} P^\nu {\tilde{\cal F}}^\lambda 
+ \frac{1}{2} q m {\tilde{\cal F}}_\mu = 0, \label{E}
\end{equation}
and 
$(P^2 + m^2) {\cal F}_\mu = (P^2 + m^2) {\tilde{\cal F}}_\mu = 0$.
Due to the linear equations (\ref{E}) we have also 
$P_\mu {\cal F}^\mu = P_\mu {\tilde{\cal F}}^\mu = 0$
and
$(P^2 + \frac{1}{4} q^2 m^2) {\cal F}_\mu 
= (P^2 + \frac{1}{4} q^2 m^2) {\tilde{\cal F}}_\mu = 0$.
Therefore, the quantum constraints are consistent, and so, 
have nontrivial solutions if and only if $|q| = 2$. 
We have arrived at the same quantization condition 
which was obtained in the classical theory.

\vskip2mm
Putting $q = \epsilon 2$, $\epsilon = +$ or $-$, we finally see that 
the field ${\cal F}^\mu$ can be identified with the topologically
massive vector U(1) gauge field ${\cal F}_\epsilon^\mu$, whereas the
field ${\tilde{\cal F}}^\mu$ coincides with ${\cal F}^\mu_{-\epsilon}$.
This gives us the desirable $P$,$T$-invariant system \cite{top}.
Let us note here that the latter can be reformulated in terms
of the gauge fields through the duality relation 
$\varepsilon_{\mu\nu\lambda} {\cal F}_\epsilon^\lambda 
= F^\epsilon_{\mu\nu}
= \partial_\mu A^\epsilon_\nu - \partial_\nu A^\epsilon_\mu$.
In this case the corresponding basic equations are 
of the second order, and can thus be compared with 
equations of motion for another $P$ and $T$ conserving system 
-- gauge-non-invariant massive model. This one, the three-dimensional 
Proca theory, describes causally propagating massive field excitations 
of spin polarizations $+1$ and $-1$. 
So, the {\it kinematical} contents of the gauge-invariant and 
non-invariant cases are identical \cite{top,Bin}. 
However, our pseudoclassical model has led exactly to topologically 
massive gauge fields. The difference between these systems may
appear {\it dynamically}, when the vector fields interact with  
matter fields \cite{Proca}.

\vskip2mm
If we choose another ordering prescription for the quantum counterpart
of the constraint function $\chi$, we would have the same operator 
but with the constant term $-3/2$ changed for $\alpha - 3/2$, where the
constant $\alpha$ specifies the ordering \cite{CPV}. As a result, we would
find that for $\alpha \neq 0, +3/2, -3/2$ under appropriate choice of
the parameter $q$ (note in this case $|q| \neq 2$) we have as a solution
of the quantum constraints only one field ${\cal F}_-^\mu$ or
${\cal F}_+^\mu$ satisfying the corresponding linear differential
equation. This would lead to the violation of the 
$P$ and $T$ symmetries at the quantum level.
For $\alpha = +3/2$ (or $ q = 0$) or $\alpha = -3/2$ the physical states 
are respectively described by one scalar field $f(x)$ or $\tilde{f}(x)$,
and for both these cases the discrete symmetries are broken. 

\vskip2mm
We see that the same values of the parameter $q$, $q = \pm 2$, which 
we have separated classically, turn out to be also special quantum
mechanically: for these the number of physical states is maximal,
so that the maximal global symmetry group can be realized on 
the physical state space, and only at $q = \pm 2$ parity and 
time-reversal symmetries are conserved. This result indicates 
a profound relationship of discrete and continuous global 
symmetries.

\section{Revealing dynamical (super)symmetries}

In what follows, we put $q = 2$, the case $q = -2$ can be achieved by 
obvious changes. To deal with the field system obtained upon quantization 
of the pseudoclassical model (\ref{L}), let us consider average value of 
the constraint operator $\widehat{\chi}$ over an arbitrary state, 
$\langle \widehat{\chi} \rangle = 
{\Psi}^\dagger(x) \widehat{\chi} \Psi(x)$. 
We find that on the trivial equations of motion for unphysical
scalar fields, $f(x) = \tilde{f}(x) = 0$, the space-time integral of
this quantity, modified on account of indefinite metric of the state
space, coincides with the action ${\cal A}$ for $P$,$T$-invariant 
system of topologically massive vector U(1) gauge fields. 
Switching to convenient matrix notations with 
Pauli matrices we can write this as follows:
\[
\int d^3x \langle\langle \widehat{\chi} \rangle\rangle 
= {\cal A} 
= \int d^3 x \Phi^\dagger(x) \left( PJ\otimes 1 
+ m\cdot 1\otimes \sigma_3 \right) \Phi(x),
\]
where we have introduced a doublet of vector fields,
$\Phi = ({\cal F}_+,{\cal F}_-)$ 
in transposed form, redefined there the scalar product to
$\langle\langle . \rangle\rangle = \Phi^\dagger . \Phi$ 
and used generators 
$(J_\mu)^\alpha{}_\beta = -i{\varepsilon^\alpha}{}_{\mu\beta}$ 
in the vector representation of the 3D Lorentz group,
$[ J_\mu,J_\nu ] = - i\varepsilon_{\mu\nu\lambda}J^\lambda$,
$J_\mu J^\mu=-2$. 
The procedure we have implemented in order to get the field action 
${\cal A}$ is reminiscent of the idea suggested in \cite{S} for 
constructing a string field theory action and subsequently
developed in \cite{W}.

\vskip2mm
Continuous global symmetries of this quantum system are 
generated by average values of the quantum counterpart 
of the third order nilpotent integrals of motion taking 
place at $q=2$. 
They are given by the expression
${\cal Q}^\pm 
= -\frac{1}{2}\langle\langle \widehat{\cal B}^\mp_+ \rangle\rangle
= \Phi^\dagger(x) Q^\pm \Phi(x)$,
where the quantum mechanical nilpotent operators
\[
Q^\pm = \frac{1}{4i} J^2_\pm \otimes \sigma_\pm
\]
realize mutual transformation of the physical states of spin
$+1$ and $-1$. 
Here we use the notation
$\sigma_\pm = \sigma_1 \pm i\sigma_2$
and $J_\pm = J^{(1)} \pm iJ^{(2)}$,
$J^{(\alpha)} = J^\mu e^{(\alpha)}_\mu$.  
Commutation relation of these operators is
\[ 
[ Q^+,Q^- ]_- = \frac{1}{2} ( S - \Pi ),
\] 
where $S = J^{(0)}\otimes 1$ 
and $\Pi = J^{(0)} J^{(0)} \otimes \sigma_3$.
$S$ is the spin operator corresponding to the average 
value of the quantum counterpart of the integral
${\cal S}$, and $\Pi$ is the operator associated with 
the projector onto the physical spin $\pm{1}$ states,
$(1 + 2\widehat{\cal N}_0)\widehat{\cal N}_\perp 
(2-\widehat{\cal N}_\perp)$.  
We have also
\[ 
[ Q^\pm,S ]_- = \pm 2 Q^\pm, \qquad 
[ Q^\pm,\Pi ]_- = \mp 2 Q^\pm.
\]
The operators $Q^\pm$ reproduce the algebra of the quantum
mechanical counterpart of the integrals ${\cal B}^\mp_+$.
When considering the following linear combinations 
\[
Q_0 = \frac{1}{4} (S - \Pi), \qquad 
Q_1 = \frac{1}{2} (Q^+ + Q^-), \qquad 
Q_2 = \frac{i}{2} (Q^+ - Q^-),
\]
we obtain that the quantum physical operators $Q_\alpha$, 
$\alpha = 0,1,2$, form $su(1,1)$ algebra,
\[
[ Q_\alpha,Q_\beta ]_- = -i \varepsilon_{\alpha\beta\gamma} Q^\gamma.
\] 
The generators $Q_\alpha$ and the Casimir operator 
$C = Q_\alpha \eta^{\alpha\beta} Q_\beta 
= \frac{3}{8} (S\Pi - \Pi^2)$ 
of this algebra form also $s(2,1)$ superalgebra 
\cite{CrRit}, 
\[
[ Q_\alpha,Q_\beta ]_+ = \eta_{\alpha\beta} \frac{2}{3} C, 
\qquad
[ Q_\alpha,C ]_- = 0.
\]   
The Casimir operator takes the value $C=-3/4$ on the physical 
subspace given by two square-integrable transversal vector fields
${\cal F}^\mu_+$, ${\cal F}^\mu_-$ carrying spins $-1$ and $+1$.

\vskip2mm
We have thus revealed hidden SU(1,1) symmetry and S(2,1) supersymmetry 
of the $P$,$T$-invariant quantum system of topologically massive vector 
U(1) gauge fields. This hidden dynamical symmetry leads to a non-standard 
super-extension of the (2+1)-dimensional Poincar\'e group. To see this, one 
needs a covariant form of the above algebra relations \cite{we}. Actually, 
the quantities $S$ and $\Pi$, as well as their combination $Q_0$, are 
expressed as covariant (scalar) operators, whereas $Q_i$ are given in 
terms of non-covariant quantities $J^{(i)}$, $i=1,2$. In the covariant 
form we found, the hidden (super)symmetry generators are presented by 
a rank-$2$ symmetric tensor, 
$X_{\mu\nu} = i e^{(0)}_\mu e^{(0)}_\nu Q_0 + X^\perp_{\mu\nu}$, 
provided that its traceless and transversal part 
$X^\perp_{\mu\nu}$
interchanges spins $+1$ and $-1$ \cite{we}.
Together with the energy-momentum $P_\mu$ and the 
total angular momentum 
$M_\mu = 
-\varepsilon_{\mu\nu\lambda}x^\nu P^\lambda 
\cdot 1 \otimes 1 + J_\mu \otimes 1$
operators, they complete the set of generators of the 
superextended Poincar\'e group $\rm ISO(2,1|2,1)$. 
The Casimir operators of this supergroup 
are $P^2$ and the {\it superspin} 
$\Sigma = \frac{1}{2}(S + \Pi)$.
The eigenvalues of the superspin
$\Sigma$ are given by the set of numbers $(-1,0,0,0,0,1)$. 
Expressing the operator $C$ through the superspin as
$C = \frac{3}{4}(\Sigma^2 - J^{(0)}J^{(0)}\otimes 1)$, 
we gain that the physical states are the eigenstates of the superspin
operator with zero eigenvalue. Hence, the one particle states of the
quantum $P$,$T$-invariant system of topologically massive vector
U(1) gauge fields realize an irreducible representation of the supergroup
$\rm ISO(2,1|2,1)$ labelled by the zero eigenvalue of the superspin.
Similar properties have been elucidated for the double fermion system 
\cite{GPS}, which is also considered to be relevant to high-temperature 
superconductivity \cite{T-inv-mod}.

\section{Concluding remarks}

In conclusion, with the help of the proposed pseudoclassical model 
(\ref{L}) we have seen that the $P$,$T$-invariant system of topologically 
massive vector U(1) gauge fields has a rich set of hidden symmetries. It 
seems very likely that namely such `internal properties' are responsible 
for critical phenomena in the corresponding models. Of course, it is not 
surprising too much, since the 3D topological mass terms \cite{top} are 
originated from the $\theta$-vacuum of 4D physics \cite{theta}, and reflect 
its rich and complicated structure. The pseudoclassical model itself turned 
out to be very interesting, having revealed the quantization of the parameter 
$q$ and nontrivial `superposition' of the discrete ($P$ and $T$) and 
continuous (hidden SU(1,1) and S(2,1)) (super)symmetries. A principal 
question to be answered concerning these results is to understand the 
sense of quantum field analog of the hidden (super)symmetry generators. 
It is quite natural to expect that they should be generators of the 
corresponding field symmetry transformations \cite{we}. One might further 
study the situation with $P,T$-invariant matter coupling and investigate 
what happens with the revealed hidden (super)symmetries.

\vskip1cm
\small
{\bf Acknowledgements}
\vskip3mm
The work of Kh.N. has been supported by the Alexander von Humboldt
Fellowship and by the European Commission TMR programme 
ERBFMRX--CT96--0045 and ERBFMRX--CT96--0090.
M.P. thanks Prof. H. P. Nilles and TU M\"unchen, where a part 
of this work has been realized, for kind hospitality.


\newpage

\begin{thebibliography}{**}

\bibitem{T}
M.B. Green, J.H. Schwarz, and E. Witten,
Superstring Theory (Cambridge University Press, Cambridge, 1987); \\
J. Polchinski, TASI Lectures on D-Branes, NSF-ITP-96-145, hep-th/9611050.

\bibitem{pm}
J.L. Martin, Proc. Roy. Soc. A 251 (1959) 536; \\
F.A. Berezin and M.S. Marinov, 
JETP Lett. 21 (1975) 678; Ann. Phys. (NY) 104 (1977) 336; \\
L. Brink, S. Deser, B. Zumino, P. Di Vecchia and P. Howe,
Phys. Lett. 64 B (1976) 435; \\
L. Brink, P. Di Vecchia and P.S. Howe,
Nucl. Phys. B 118 (1977) 76; \\
R. Casalbuoni, Nuovo Cim. 33 A (1976) 369; \\
A. Barducci, R. Casalbuoni and L. Lusanna,
Nuovo Cim. 35 A (1976) 377; \\
C.A. Galvao and C. Teitelboim, 
J. Math. Phys. 21 (1980) 1863; \\
M. Henneaux and C. Teitelboim, 
Ann. Phys. (NY) 143 (1982) 127.  

\bibitem{Ber}
F.A. Berezin, The Method of Second Quantization
(Academic Press, New York, 1966).

\bibitem{GS}
M.B. Green and J.H. Schwarz, 
Phys. Lett. B 136 (1984) 367; \\
see also I.A. Bandos, D. Sorokin, M. Tonin, 
P. Pasti and D.V. Volkov,
Nucl. Phys. B 446 (1995) 79 and refs. therein.

\bibitem{BS}
L. Brink and J.H. Schwarz, 
Phys. Lett. 100 B (1981) 310.

\bibitem{NP1}
Kh.S. Nirov and M.S. Plyushchay, 
Phys. Lett. B 405 (1997) 114 [hep-th/9707070].

\bibitem{top}
R. Jackiw and S. Templeton, 
Phys. Rev. D 23 (1981) 2291; \\ 
J. Schonfeld, Nucl. Phys. B 185 (1981) 157; \\ 
W. Siegel, Nucl. Phys. B 156 (1979) 135; \\ 
S. Deser, R. Jackiw and S. Templeton, 
Phys. Rev. Lett. 48 (1982) 975; 
Ann. Phys. (NY) 140 (1982) 372.

\bibitem{hall}
S.M. Girvin and A.H. MacDonald, 
Phys. Rev. Lett. 58 (1987) 1252; \\
S. Girvin and R. Prange (eds), The Quantum Hall Effect 
(Springer-Verlag, Berlin, 1987); \\
Z. Zhang, T. Hansson and S. Kivelson, 
Phys. Rev. Lett. 62 (1989) 82 (E), 980; \\ 
N. Read, Phys. Rev. Lett. 62 (1989) 86.

\bibitem{T-sc}
A. Polyakov, Mod. Phys. Lett. A 3 (1988) 325; \\ 
Y.-H. Chen, F. Wilczek, E. Witten and B. Halperin,
Int. J. Mod. Phys. B 3 (1989) 1001. 

\bibitem{GN-AI}
M. Gell-Mann and Y. Ne'eman, The Eightfold Way
(W. A. Benjamin, New York, 1964); \\
A. Arima and F. Iachello, Ann. Phys. (NY) 99 (1976) 253.

\bibitem{R}
P. Ramond, Phys. Rev. D 3 (1971) 2415.

\bibitem{GL}
Yu. Gol'fand and E. Lichtman, JETP Lett. 13 (1971) 323. 

\bibitem{dual}
P.K. Townsend, K. Pilch and P. van Nieuwenhuizen,
Phys. Lett. 136 B (1984) 38; \\
S. Deser and R. Jackiw, Phys. Lett. 139 B (1984) 371.

\bibitem{T-inv}
X.G. Wen and A. Zee, 
Phys. Rev. Lett. 62 (1989) 2873.

\bibitem{exp}
See, for example,  
A. Mathai, Y. Gim, R.C. Black, A. Amar and F.C. Wellstood,
Phys. Rev. Lett. 74 (1995) 4523.

\bibitem{cancel}
J. Lykken, J. Sonnenschein and N. Weiss, 
Phys. Rev. D 42 (1990) 2161; \\
N.E. Mavromatos, 
University of Oxford Preprint OUT--90--04P (1990).

\bibitem{T-inv-mod}
A. Kovner and B. Rosenstein, Phys. Rev. B 42 (1990) 4748; \\
G.W. Semenoff and N. Weiss, Phys. Lett. B 250 (1990) 117; \\
N. Dorey and N.E. Mavromatos, Phys. Lett. B 250 (1990) 107;
Nucl. Phys. B 386 (1992) 614.

\bibitem{PR}
M.S. Plyushchay and A.V. Razumov, 
Int. J. Mod. Phys. A 11 (1996) 1427 [hep-th/9306017].

\bibitem{Bin}
B. Binegar, J. Math. Phys. 23 (1982) 1511.

\bibitem{Proca}
In this respect, it is of interest to investigate 
quantum symmetries of the three-dimensional Proca 
model; see, for example, 
J. Beckers, N. Debergh and A.G. Nikitin, 
Fortsch. Phys. 43 (1995) 81, 
where hidden parasupersymmetries of the 
four-dimensional Proca theory were analyzed. 

\bibitem{CPV}
J.L. Cort\'es, M.S. Plyushchay and L. Vel\'azquez, 
Phys. Lett. B 306 (1993) 34. 

\bibitem{S}
W. Siegel, Phys. Lett. 149 B (1984) 157, 162,
corrected in Phys. Lett. 151 B (1985) 391, 396. 

\bibitem{W}
E. Witten, Nucl. Phys. B 268 (1986) 253, 
ibid. B 276 (1986) 291. 

\bibitem{CrRit}
M. de Crombrugghe and V. Rittenberg, 
Ann. Phys. (NY) 151 (1983) 99.

\bibitem{we}
Kh.S. Nirov and M.S. Plyushchay, 
a report with details of this work is in preparation. 

\bibitem{GPS}
G. Grignani, M. Plyushchay and P. Sodano, 
Nucl. Phys. B 464 (1996) 189.

\bibitem{theta}
R. Jackiw and C. Rebbi, 
Phys. Rev. Lett. 37 (1976) 172;
R. Jackiw, Rev. Mod. Phys. 52 (1980) 661;
S. Deser, M. Duff and C. Isham, 
Phys. Lett. 93 B (1980) 419.

\end{thebibliography}
\end{document}